\documentclass[12pt]{article}
\usepackage{amssymb,amsmath}

\newcommand{\dd}{\mbox{\rm d}}

\newcommand{\gam}{\gamma}

\newcommand{\tl}{\tilde}

\newcommand{\DD}{\mbox{\rm D}}

\newcommand{\p}{\partial}

\newcommand{\be}{\begin{equation}}
\newcommand{\bear}{\begin{eqnarray}}
\newcommand{\ear}{\end{eqnarray}}
\newcommand{\ee}{\end{equation}}
\newcommand{\lbl}{\label}
\newcommand{\bi}{\bibitem}
\newcommand{\ci}{\cite}

\newcommand{\vs}{\vspace}

\begin{document}

\

\baselineskip .47cm 

%\vs{10mm}

\begin{center}

{\LARGE \bf On the Relativity in Configuration Space:
A Renewed Physics In Sight}\footnote{ Work supported by the Slovenian Research
Agency.}

\vs{3mm}

Matej Pav\v si\v c

Jo\v zef Stefan Institute, Jamova 39,
1000 Ljubljana, Slovenia

e-mail: matej.pavsic@ijs.si

\vs{6mm}

{\bf Abstract}

\end{center}

\baselineskip .4cm

%\vs{1mm}

{\small

The idea that possible configurations of a physical system can be represented
as points in a multidimensional configuration space ${\cal C}$
is explored. The notion of spacetime, without
${\cal C}$, does not exist in this theory. Spacetime  is associated with
the degrees of freedom of a chosen single particle within a
considered configuration, and is thus a subspace of ${\cal C}$.
Finite dimensional configuration spaces of point particles, and
infinite dimensional configuration spaces of branes are considered.
Multidimensionality of a configuration space has for a consequence
the existence of extra interactions, besides the 4D gravity, both at
macroscopic and microscopic scales.


%\vs{3mm}

\baselineskip .55cm 

\section{Introduction}

Occasionally a fresh look at a well established theory may bring surprises.
In this paper I will discuss and further develop an approach to description of
many particle and extended systems which was
considered in ref.\,\ci{PavsicBook,PavsicBled}. Related work has been
done in refs.\,\ci{Barbour}--\ci{AuriliaFuzzy}.
Paraphrasing
Feynman\footnote{``...every theoretical physicist
who is any good knows six or seven different theoretical
representations of exactly the same physics\,\ci{Feynman}."}
a full understanding of one and the same physics
requires at least six or seven different representation. According
to that famous remark it should be thus desirable to discover
some alternative ways of describing a system of point particles or
a system of strings and branes. The latter objects are amongst the most
active topics of current research in fundamental theoretical physics.

Usually, all those objects are considered to live in a background spacetime.
In spacetime we thus have  ``matter" consisting of all sorts of
physical objects, such as branes of various dimensions, including
point particles and strings.
But we can look at the situation from another angle. We can consider spacetime
as a space of all possible positions of a chosen single particle
(a ``test particle")
while keeping fixed positions of all other particles. In other words,
spacetime can be considered as the {\it
configuration space of a single point particle} relative to the (assumed)
fixed position of the remaining particles within the `full' configuration.
This is a deviation from the standard physical thought, where there are
subtle differences between how the configuration and the usual space are
to be treated.

The configuration space of a single point particle is just a start in a
construction of physical theories. We can include other particles and extended
objects into the description as well, and consider a {\it multidimensional
configuration space of a system of particles or extended objects}.
Usually, an action for a system of (free) point particles or,
in general, branes, is written as the sum of one particle (brane)
action. But there is a fascinating
possibility to go beyond the existing physics which takes place
in spacetime. {\it We can formulate physics in configuration space and
take the latter space as the arena for physics}. Similar approaches
were previously proposed within the context of an infinite dimensional
space of branes, called ${\cal M}$-space \ci{PavsicBook}, and within
the context of 16-dimensional space of points, areas and volumes,
called {\it Clifford space}, or shortly, $C$-space
\ci{CastroPavsicReview}--\ci{PavsicKaluza},\ci{PavsicBook}.
Both spaces are particular cases of configuration
space. The former one is an infinite dimensional space of all possible
brane (or many brane) configurations, while the latter one is a space
of all possible polyvectors (superpositions of $r$-vectors)
associated with extended objects. Clifford space $C$ is a manifold whose
tangent space at any point is a Clifford algebra. $C$ can be flat, but in
general it
 can be curved. Flat Clifford space ($C$-space) provides a possible
generalization of special relativity, whilst curved $C$-space provides
a generalization of general relativity. Metric of curved 16-dimensional
$C$-space can describe, \` a la Kaluza-Klein, the ordinary 4-dimensional
gravitational field and gauge fields due to other interactions 
\ci{PavsicKaluza}.

\section{System of point particles}

A system of free relativistic point particles in $N$-dimensional
spacetime can be described by the action
\be
    I[X^\mu] =  \sum_{i=1}^n m_i \int \dd \tau_i 
    ({\dot X}_i^\mu {\dot X}_{i \mu})^{1/2} ~, ~~~~~~i = 1,2,...,n;~~
    \mu = 0,1,2, N-1
\lbl{2.1}
\ee
which is the sum of single particle actions. Here $n$ denotes the number
of particles and $m_i$  the i-th particle mass. Equations of motion derived
from (\ref{2.1}) are
\be
     m_i \frac{\dd}{\dd \tau_i} \left ( 
   \frac{{\dot X}_i^\mu}{\sqrt{{\dot X}_i^2}} \right ) = 0
,
\lbl{2.2}
\ee
where ${\dot X}_i^2 \equiv {\dot X}_i^\nu {\dot X}_{i \nu} =
{\dot X}_i^2 (\tau_i)$.
Each particle within the system moves as free particle, its mass being
$m_i$.

Let us now consider the following action:
\be
    I[X_i^\mu] = M \int \dd \tau ({\dot X}_1^\mu {\dot X}_1^\nu \eta_{\mu \nu}
    +  {\dot X}_2^\mu {\dot X}_2^\nu \eta_{\mu \nu}  +  
    {\dot X}_3^\mu {\dot X}_3^\nu \eta_{\mu \nu} + ... + 
    {\dot X}_n^\mu {\dot X}_n^\nu \eta_{\mu \nu} )^{1/2} ,
\lbl{2.1a}
\ee
where $M$ is a constant and $\tau$ and arbitrary monotonically increasing
parameter.

Writing ${\dot X}_i^\mu \equiv {\dot X}^{\mu i} \equiv {\dot X}^M,~
M\equiv (\mu i)$, $\mu = 0,1,2,3; ~ i=1,2,...,n$, then the action
(\ref{2.1a}) becomes
\be
     I[X^M] = M \int \dd \tau ({\dot X}^M {\dot X}^N \eta_{MN} )^{1/2} ,
\lbl{2.1b}
\ee
where
\be
    \eta_{MN} \equiv \eta_{(\mu i)(\nu j)} = \eta_{\mu \nu} \delta_{ij} .
\lbl{2.1c}
\ee
Eq.\,(\ref{2.1b}) is the minimal length action in flat configuration space
${\cal C}$ spanned by a system of {\it free} point particles, $\eta_{MN}$
being the diagonal metric of ${\cal C}$. The corresponding equations of 
motion are
\be
    M \frac{\dd}{\dd \tau} \left ( 
    \frac{{\dot X}^M}{({\dot X}^N {\dot X}_N)^{1/2}} \right ) = 0 .
\lbl{2.1d}
\ee
Action (\ref{2.1b}) is invariant with respect to reparametrizations of
$\tau$. Taking a gauge in which ${\dot X}^N {\dot X}_N \equiv
{\dot X}^{\mu i} {\dot X}^{\nu j} \eta_{(\mu i)(\nu j)}$ is constant,
we have ${\ddot X}^M \equiv {\ddot X}^{\mu i} = 0$, which implies that
${\dot X}^{\mu i}$ is constant. As a consequence, also the quadratic form
${\dot X}^{\mu i} {\dot X}^{\nu i} \eta_{\mu \nu }$ for a single particle,
labeled by $i$, is constant, which is just a gauge fixing condition
for the ordinary equation of motion (\ref{2.2}). Such choice of gauge in
eq.\,(\ref{2.2}) also gives ${\ddot X}^{\mu i} = 0$.

We thus see that the second action (\ref{2.1b}) gives the same equations
of motion for the $i$-th particle as the usual action (\ref{2.1}).
For {\it free particles} we may use either the usual action which is the
sum of point particle actions, or we may use the action (\ref{2.1b}) which
is proportional to the length in configuration space ${\cal C}$.
{\it The difference occurs when we consider interactions}. This will be
explored in next sections. The form of the action (\ref{2.1b}) suggests
that we have now the theory of relativity in configuration space, quite
analogous to the theory of relativity in spacetime.

In the case of {\it flat configuration space} ${\cal C}$ the law of motion is
given by eq.\,(\ref{2.1d}) which says that a configuration, represented
by a point in ${\cal C}$, traces a flat world line in ${\cal C}$. This means
that in spacetime, every particle traces a flat worldline.

Just as the ordinary Lorentz transformations preserve the quadratic
form
\be
      (x'^\mu - {x'_0}^\mu)(x'^\nu - {x'_0}^\nu) \eta_{\mu \nu} =
      (x^\mu - {x_0}^\mu)(x^\nu - {x_0}^\nu) \eta_{\mu \nu}
\lbl{2.1e}
\ee
so in the configuration space we have analogous transformations which preserve
\be
      (x'^M - {x'_0}^M)(x'^N - {x'_0}^N) \eta_{MN} =
      (x^M - x_0^M)(x^N - x_0^N) \eta_{MN} .
\lbl{2.1f}
\ee
Thus $(x'^M - {x'_0}^M) = {L^M}_J (x^J - x_o^J)$,
where the transformation matrix has to satysfy
${L^M}_J {L^n}_K \eta_{MN} = \eta_{JK}$.
Here $x^M - x_0^M \equiv x^{\mu i} - x_0^{\mu i}$ is the difference of
coordinates of two configurations. The group of Lorentz transformations
in multidimensional space ${\cal C}$
contains a subgroup of the ordinary Lorentz transformations that preserve
the 4-dimensional quadratic form (\ref{2.1e}). According to this picture
Lorentz transformations in spacetime are just particular transformations,
whereas in general we have Lorentz transformations in ${\cal C}$.

\section{Point particle in curved configuration space }

We will assume that, in general, a space  ${\cal C}$ need not be
flat, but may have non vanishing curvature.
Instead of the flat space action
(\ref{2.1b}) we have now the action in the presence of a background
metric field $G_{MN} (X)$ which depends on points $x^M$ of ${\cal C}$:
\be
   I[X^M] = M \int \dd \tau \left ({\dot X}^M (\tau) {\dot X}^N (\tau) G_{MN}
   \right )^{1/1} .
\lbl{3.1}
\ee
From the point of view of the underlying 4-dimensional spacetime
$M_4$ (which is a subspace of ${\cal C}$) we have a system of worldlines,
described by functions $X^M (\tau)\equiv X^{i \mu} (\tau)$. If there are
no interactions between the particles, then the worldlines are straight
lines in $M_4$; a point in  ${\cal C}$ traces a straight line.
Configuration space  ${\cal C}$ is then flat and its metric
$G_{MN}$ is that of a flat space, considered in previous section.
One can choose a coordinate system in
which $G_{MN} \equiv G_{(i\mu)(j \nu}$ is diagonal metric at all points
of  ${\cal C}$:
\be
    G_{MN} = \eta_{M N} .
\lbl{3.2}
\ee
In general this need not be the case: The metric $G_{MN}$ of the configuration
space space can have non vanishing off diagonal terms that cannot be
transformed away by a choice of coordinates.

The off diagonal terms $G_{(\mu i)(\nu j)},~i\neq j$ are responsible
for the {\it interactions between the particles} which according to this
novel theory exist besides the ordinary gravitational interaction
incorporated in the metric $G_{(\mu i) (\nu i)} \equiv g_{\mu \nu}$.
In ref.\,\ci{PavsicBled} we provided arguments that such approach might explain on the
cosmological and astropphysical scales the puzzles of ``dark matter" or
``missing mass", and on the microscopic scale the existence
of electroweak and color interactions.

From the action (\ref{3.1}) we obtain the equation of geodesic in the
presence of a metric $G_{MN}$:
\be
     \sqrt{{\dot X}^2}\, \frac{\dd}{\dd \tau} \left (\frac{\dot X^M}{\sqrt{\dot
     X^2 }} \right ) + \Gamma_{JK}^M \dot X^J \dot X^K = 0
.
 \lbl{3.4}
 \ee
The configuration space metric $G_{MN}$ causes that a worldline $X^M (\tau)$
in general is not a straight line in ${\cal C}$ and thus also the worldlines
of particles are not straight lines in $M_4$. The term with connection 
$\Gamma$ occurring in the geodesic equation (\ref{3.4}) has a role of
a ``force" in ${\cal C}$, and manifests itself in $M_4$ as the interactions
between the particles.

In this approach a configuration of a system of particles is
considered as a whole, namely as a point in configuration space, which is
the space of all possible configurations. We postulate that the latter
space is endowed with metric, connection, and curvature. Metric $G_{MN}$
should be considered as a dynamical quantity, its kinetic term
being given by the Einstein-Hilbert action in ${\cal C}$ whose dimension
is $D = N \times n$:
\be
     I[G_{MN}] = \frac{1}{16 \pi} \int \dd^D x \, \sqrt{|G|} \, \cal R
 ,
\lbl{3.5}
\ee
where $\cal R$ is the Ricci scalar in ${\cal C}$, and
$G \equiv {\rm det} \, G_{MN}$. The total action is the sum
of $I[X^M]$ and $I[G_{MN}]$.
It is invariant under general coordinate
transformations in ${\cal C}$.

We will leave aside a detailed study of solutions to such a system.
For the purpose of the present paper it suffices if we make a plausible
assumption that within a set of solutions there exist solutions
with isometries. Let us therefore suppose that as a solution to our
dynamical system there can exist a space ${\cal C}$ which
admits $K$ Killing vector fields ${k^\alpha}_M$, $\alpha = 1,2,...,K$,
satisfying
$\DD_M {k^\alpha}_N + \DD_N {k^\alpha}_M = 0$
,
where the covariant derivative $D_M$ is defined with respect to the
metric $G_{MN}$ of configuration space ${\cal C}$.

Let us split the indices according to $M=(\mu 1,\bar M)$, where $\mu 1 \equiv
\mu = 0,1,2,3$ are indices of coordinates of a chosen single particle,
say a particle No. 1 (i.e., with $i=1$), whilst $\bar M$ are indices
of coordinates of all the remaining particles within the system.
Then the metric can be written as
\be
     G_{MN} =  \begin{pmatrix}
          G_{\mu \nu} - \phi^{\bar M \bar N} {k_ \alpha}_{\bar M}
          {k_\beta}_{\bar N}{A^\alpha}_\mu {A^\beta}_\nu~, &
          {k_\alpha}_{\bar M} {A^\alpha}_\nu \\
          {k_\alpha}_{\bar N} {A^\alpha}_\mu~, & \phi_{\bar M \bar N}
       \end{pmatrix}
,
 \lbl{3.7}
 \ee
where $ \phi^{\bar M \bar N}
$ is the inverse of $\phi_{\bar M \bar N}$ in the ``internal space", and
where a coordinate system in which ${k_\alpha}^\mu =0$ and
${k_\alpha}^{\bar M} \neq 0$ has been used.

Using metric (\ref{3.7}), a  quadratic form in ${\cal C}$ can be split
into a 4-dimensional part plus the part due to the remaining dimensions
of ${\cal C}$. We will now apply this to the action (\ref{3.1}).
Since it is a reparametrization invariant action, there exists a
constraint
\be
    P^M P^N G_{MN} - M^2 = 0
,
\lbl{3.8}
\ee
where
\be
    P^M = \frac{M \dot X^M}{ (\dot X^N \dot X_N)^{1/2}}
\lbl{3.8a}
\ee
are contravariant components of the momentum conjugate to coordinates
$X^M$. Inserting eq.\,(\ref{3.7}) into eq.\,(\ref{3.8}) we have
\be
   M^2 = g_{\mu \nu} p^\mu p^\nu + \phi^{\bar M \bar N} 
   p_{\bar M} p_{\bar N}
,
\lbl{3.9}
\ee
where $g_{\mu \nu} =  G_{\mu \nu} - \phi^{\bar M \bar N}
{k_ \alpha}_{\bar M} {k_\beta}_{\bar N}{A^\alpha}_\mu {A^\beta}_\nu$.
From eq.\,(\ref{3.9}) we find that the 4-dimensional mass is
\be
   m \equiv \sqrt{g_{\mu \nu} p^\mu p^\nu} = \sqrt{M^2 -
    \phi^{\bar M \bar N}  p_{\bar M} p_{\bar N}}
.
\lbl{3.10}
\ee
According to the latter relation, a mass $m$ of a single particle, defined
by means of the 4-dimensional momentum quadratic form, depends on the
momenta $p_{\bar M} \equiv p_{i \mu},~i\neq 1$, of all the remaining particles
within the considered system, which could be the entire universe.
This is reminiscent of Mach's principle.

Now let us investigate whether the 4-dimensional mass $m$ can be a constant
of motion. Obviously, the configuration space mass $M$ is constant, whatever
the metric $G_{MN}$. In a trivial case, if $G_{MN} = \eta_{MN}$ at all points
of ${\cal C}$, then $m$ is a constant of motion. We will show that
$m$ can be a constant of motion in the case of a more general metric as well,
if the space ${\cal C}$ admits suitable isometries. 

The metric $\phi^{{\bar M}{\bar N}}$ of the internal space
can be rewritten in terms of a metric
$\varphi^{\alpha \beta}$ in the space of isometries:
\be
\phi ^{\bar M\bar N}  = \varphi ^{\alpha \beta } k_\alpha  ^{\bar M} k_\beta  ^{\bar N}
+ \phi_{\rm extra}^{{\bar M}{\bar N}} .
\lbl{3.11}
\ee
Here $\phi_{\rm extra}^{{\bar M}{\bar N}}$ are additional terms due to the directions
that are orthogonal to isometries. For particular
internal spaces ${\bar {\cal C}}$, those additional terms may vanish. Let us
assume that this is the case.

Inserting eq.\,(\ref{3.11}) into eq.\,(\ref{3.10})we have
\be
    m = (M^2 - \varphi^{\alpha \beta}p_\alpha p_\beta)^{1/2}
,
\lbl{3.14}
\ee
where $p_\alpha \equiv {k_\alpha}^{\bar M} p_{\bar M}$  is a {\it constant
of motion} due to the $\alpha$-th isometry\footnote{One possibility is to
choose isometries ${k_\alpha}^M$ in the full configuration space ${\cal C}$.
Then the projections $p_\alpha$ of momenta $P_M$ onto the Killing
vectors  ${k_\alpha}^M$ are constants of motion.
Another possibility is to consider the isometries of the `internal'
subspace of ${\cal C}$. Then $p_\alpha = {k_\alpha}^{\bar M} p_{\bar M}$
are not constants in general, whereas the quadratic form
$\varphi^{\alpha \beta}p_\alpha p_\beta$ can be a constant.}. 
{\it So also 4-dimensional
mass $m$ is a constant of motion} in this particular case of appropriate
isometries.

A consequence of the latter property is that a particle accelerated by means
of ``forces" due to the metric (\ref{3.7}) of a configuration space
${\cal C}$ which admits the above isometries cannot exceed the speed
of light in 4-dimensional spacetime.

This can be shown by considering the momentum (\ref{3.8a}) and the relation
\be
   m = M \left ( \frac{\dot X^\mu \dot X^\nu g_{\mu \nu}}
    {\dot X^M \dot X^N G_{MN}} \right )^{1/2}
\lbl{3.15}
\ee
which is a consequence of
\be
   \dot X^M \dot X^N G_{MN} = \dot X^\mu \dot X^\nu g_{\mu \nu}
   + \dot X_{\bar M} \dot X_{\bar N} \phi^{\bar M \bar N}
\lbl{3.16}
\ee
and eq.\,(\ref{3.9}). Here
\be
    g_{\mu \nu} =  G_{\mu \nu} - \phi^{\bar M \bar N}
   {k_ \alpha}_{\bar M} {k_\beta}_{\bar N}{A^\alpha}_\mu {A^\beta}_\nu
.
\lbl{3.17}
\ee
is 4-dimensional metric.
Using eq.\,(\ref{3.15}), the momentum (\ref{3.8a}) can be rewritten
as
\be
   p^M = \frac{m\,\dot X^M}{(\dot X^\rho \dot X^\sigma g_{\rho \sigma})^{1/2}}
.
\lbl{3.18}
\ee
For the components $M = \mu 1 \equiv \mu =0,1,2,3$ of a single particle within
our multiparticle system we have thus
\be
 p^\mu = \frac{m \,\dot X^\mu}{(\dot X^\rho \dot X^\sigma g_{\rho \sigma})^{1/2}}
\lbl{3.19}
\ee
If $m$ is a {\it constant}, which is indeed the case in the presence of
the considered isometries, then, according to eq.\,(\ref{3.19}), the condition
for $p^\mu$ to remain real is
\be
 \dot X^\rho \dot X^\sigma g_{\rho \sigma} \ge 0 . 
\lbl{3.19a}
\ee
In other words, in spite of the fact that the particle is being accelerated
(i.e., moving along a geodesic of the configuration space ${\cal C}$),
its limiting speed in the subspace $M_4$ is the speed of light.

This is not the case in a more general configuration space, which does
not admit Killing vector fields. Then the general expression for
momentum eq.\,(\ref{3.8a}) cannot be reduced to the form (\ref{3.18})
with $m$ being a constant of motion.
{\it A prediction of this theory is thus that the speed of light in $M_4$
is the
 limiting speed for a particle which is accelerated by gauge fields
${A^\alpha}_\mu$ (including the electromagnetic field $A_\mu$) that arise
in the presence of isometries, but is not a limiting speed in a more
general case when isometries are absent.}

The above refers to the limiting speed in 4-dimensional spacetime,
which is a subspace of the configuration space ${\cal C}$. In the
latter larger space, because of the relation (\ref{3.8a}) for momentum,
it always holds
\be
    \dot X^M \dot X^N G_{MN} > 0
\lbl{3.20}
\ee
regardless of whether there are isometries or not,. That is, in
${\cal C}$ there is a limiting speed, determined by the
condition (\ref{3.20}). The latter limiting speed involves not
only four spacetime components due to a single particle, but
also components $\dot X^{\bar M}$ due to the presence of
other particles.

Let us now explicitly show that in the presence of isometries,
which imply that $\varphi^{\alpha \beta} p_\alpha p_\beta$ is
constant, conditions (\ref{3.19a}) and (\ref{3.20}) are consistent.
Indeed, from
\be
   \varphi^{\alpha \beta} p_\alpha p_\beta  = \phi^{\bar M \bar N}
   p_{\bar M} p_{\bar N} = \frac{M^2 \phi^{\bar M \bar N} \dot X_{\bar M}
   \dot X_{\bar N}}{\dot X^\mu \dot X^\nu g_{\mu \nu} +  
   \phi^{\bar M \bar N} \dot X_{\bar M}\dot X_{\bar N} }
\lbl{3.21}
\ee
in which we denote $\dot X^\mu \dot X^\nu g_{\mu \nu} \equiv X$,
$~~\phi^{\bar M \bar N} \dot X_{\bar M} \dot X_{\bar N} \equiv Y$ and
$\varphi^{\alpha \beta} p_\alpha p_\beta/M^2 \equiv C$ we have that
for a fix chosen constant $C$ there is a proportionality between $X$ and
$Y$:
\be
      Y = \frac{C}{1-C}\,X
 .
\lbl{3.23}
\ee
Therefore $X$ and $Y$ cannot change independently; if $X$ approaches
zero, also $Y$ approaches zero, and so does the sum $X+Y \equiv
G_{MN} \dot X \dot X^N$. This proves consistency of the conditions
(\ref{3.19a}) and (\ref{3.20}). However, if there are no isometries,
then $C$ is not a constant of motion, and condition (\ref{3.19a})
does not hold. However, condition (\ref{3.20}) which imposes
a restriction on velocities in configuration space remains valid.

The signature of the configuration space ${\cal C}$ is in general $(p,q)$,
and thus in ${\cal C}$ there is no separation between different regions,
that could be identified with past, present and future. And yet, if
${\cal C}$ admits isometries, as described above, then the concept of
light cone in a subspace $M_4$, with distinction between past, present
and future, makes sense. This is so, because a particle's 4-dimensional mass
$m$ is then a constant of motion, and the particle cannot pass the light barrier
during its motion in $M_4$.

The ordinary relativity in 4-dimensional spacetime is thus embedded
in the more general relativity that holds in a multidimensional space
${\cal C}$.
 
\section{Configuration space for strings and branes}

String and brane theories are very elegant and promising
in explaining the origin and interrelationship
of the fundamental interactions, including gravity\,\ci{strings,branes}.
But such theories are still far from being finished. One of the
unsettled problems is a question of the geometric principle behind the string
and brane theories\,\ci{geometric}. For a recent
serious criticism see\,\ci{Smolin}. In the following
we will consider the possibility that string/brane theories
should take into account the concept of configuration
space. 

A brane configuration
can be described by the set of functions $X^\mu (\xi^a)$, where $\xi^a$,
$a=1,2,...,n$, is a set of parameters on the brane.
We will consider a brane configuration as a point in an infinite
dimensional configuration space, called brane space ${\cal M}$.
Following refs.\,\ci{PavsicBook,PavsicBled}, we will therefore
use a condensed notation
\be
X^\mu  (\xi ^a )\, \equiv \,X^{\mu (\xi )} \, \equiv X^M .
\lbl{4.1}
\ee
We assume that the branes within classes of tangentially deformed branes 
are in principle physically distinct objects. 
All such objects are represented by different points of 
${\cal M}$-space.

Instead of one brane we can take a 1-parameter family of
branes $ X^\mu  (\tau,\xi ^a )\, \equiv \,X^{\mu (\xi )}(\tau)
 \, \equiv X^M (\tau) $, i.e., a curve (trajectory) in ${\cal M}$.
In principle every trajectory is kinematically possible.
A particular dynamical theory then selects which amongst
those kinematically possible branes and trajectories are
dynamically possible. We assume that dynamically possible
trajectories are {\it geodesics} in ${\cal M}$ determined by
the minimal length action \ci{PavsicBook}:
\be
I[X^M ] = \int {{\rm{d}}\tau } \,\,(\rho _{MN} \,\dot X^M \dot X^N )^{(1/2)} .
\lbl{4.2}
\ee
Here $\rho_{MN}$ is the metric of ${\cal M}$.

In particular, if metric is
\be
\rho _{MN} \, \equiv \,\rho _{\mu (\xi ')\nu (\xi '')} \, = 
\,\kappa \,\frac{{\sqrt {|f(\xi ')|} }}{{\sqrt {\dot X^2 \,(\xi ')} }}
\,\delta (\xi ' - \xi '')\,\eta _{\mu \nu } ,
\lbl{4.3}
\ee
where $f_{ab} \, \equiv \partial _a X^\mu  \partial _b X^\nu  \eta_{\mu \nu } $
is the induced metric on the brane,  $f \equiv \det \,f_{ab}$,
$\dot X^2  \equiv \dot X^\mu  \dot X^\nu  g_{\mu \nu } $, ($\eta_{\mu \nu}$
being the Minkowski metric of the embedding spacetime), then the equations
of motion derived from (\ref{4.2}) are precisely those of a Dirac-Nambu-Goto
brane \cite{PavsicBook}. Although we started from
a brane configuration space in which tangentially deformed branes
are considered as distinct objects, the dynamical theory, based on
the action (\ref{4.2}) and the particular choice of metric (\ref{4.3}),
has for solutions the branes which satisfy such constraints which imply that
only the transversal excitations are physical, whereas the tangential
excitations are nothing but reparametrizations of $\xi^a$ and $\tau$.
For more details see ref.\,\ci{PavsicBook}.

In this theory we assume that metric (\ref{4.3}) is just one particular
choice amongst many other possible metrics of ${\cal M}$. But
dynamically possible metrics are not arbitrary. We assume that
they must be solutions of the Einstein equations in ${\cal M}$
\cite{PavsicBook}. 

We take the brane space ${\cal M}$ as an arena for physics.
The arena itself is a part of the dynamical system,
it is not prescribed in advance.
The theory is thus background independent. It is based on the geometric
principle which has its roots in the brane space ${\cal M}$.

To sum up, the infinite dimensional brane space ${\cal M}$ has in
principle any metric that is a solution to the Einstein's equations in
${\cal M}$. For the particular diagonal metric (\ref{4.3}) we obtain the ordinary
branes, including strings. But it remains to be checked whether such
particular metric is a solution of this generalized dynamical system
at all. If not, then this would mean that the ordinary string and brane
theory is not exactly embedded into the theory based on dynamical
${\cal M}$-space. The proposed theory goes beyond that 
of the usual strings and branes. It  resolves the problem
of background independence and the geometric principle behind the
string theory. Geometric principle behind the string theory
is based on the concept of brane space ${\cal M}$, i.e., the configuration
space for branes. Occurrence of gauge and gravitational fields in
string theories is also elucidated. Such fields are due to string
configurations. They occur in the expansion of a string state
functional in terms of the Fock space basis. A novel insight is that
they occur even within the classical string theory based on the
action (\ref{4.2}) with ${\cal M}$-space metric $\rho_{MN}$,
which is dynamical and satisfies the Einstein equations in ${\cal M}$.
Multidimensionality of $\rho_{MN}$ allows for extra gauge interactions,
besides gravity. In the following we will discuss how in the infinite
dimensional space ${\cal M}$ one can factor out a finite
dimensional subspace.

\section{Finite dimensional description of extended objects}

When considering the motion of macroscopic extended objects such as planets,
we usually take into account a finite set of degrees of freedom only,
e.g., the coordinates of the center of mass, and neglect all the
remaining many degrees of freedom. Similarly, when considering, e.g.,
a closed string, we can describe it in the first approximation by
four coordinates $X^\mu$ of the center of mass.
In the next approximation we can describe
it in terms of the coordinates $X^{\mu_1 \mu_2}$ of the oriented area
enclosed by the string. If the string has finite thickness and thus
it actually is not a string but a 2-brane, then we can also consider
the corresponding volume degrees of freedom $X^{\mu_1 \mu_2 \mu_3}$.

In general, an extended object in 4-dimensional spacetime can be
described by 16 coordinates
\be
X^M \, \equiv \,X^{\mu _1 ...\mu _r } \,,\,\,\,\,\,\,\,r = 0,1,2,3,4 \,.
\lbl{5.1}
\ee
They are the projections of $r$-dimensional volumes (areas) onto the
coordinate planes, and
they denote a point in a 16-dimensional space $C$, which is a subspace
of the full infinite dimensional space ${\cal M}$, the configuration
space of the considered extended object.

Oriented $r$-volumes can be elegantly described by Clifford
algebra\,\ci{Hestenes}. Let us illustrate this on the example of
a 2-surface $\Sigma$ bounded by
a loop $B$. An infinitesimal surface element is given by the wedge
product of two infinitesimal vectors $\dd \xi_1$ and $\dd \xi_2$, expanded
in terms of the basis tangent vectors $e_a$, $a=1,2$:
\be
    \dd \Sigma = \dd \xi_1 \wedge \dd \xi_2 = \dd \xi_1^a \dd \xi_2^b
     \, e_a \wedge e_b = \frac{1}{2} \, \dd \xi^{ab} e_a \wedge e_b ,
\lbl{5.1a}
\ee
where $\dd \xi^{ab} = \dd \xi_1^a \dd \xi_2^b - \dd \xi_2^a \dd \xi_1^b$.
Inserting the relation $e^a = \p_a X^\mu \gamma_\mu$ between the
basis vector $e_a$  tangent to the surface $\Sigma$
and the basis vectors $\gamma_\mu$ of the embedding space(time),
and integrating over $\dd \Sigma$, we have
\be
    \int_{\Sigma_B} \dd \Sigma = \frac{1}{2} X^{\mu \nu}
    \gam_\mu \wedge \gam_\nu ,
\lbl{5.1b}
\ee
where
\be
      X^{\mu \nu} = \frac{1}{2} \int_{\Sigma_B} \dd \xi^{ab}
      (\p_a X^\mu \p_b X^\nu - \p_a X^\nu \p_b X^\mu) ,
\lbl{5.1c}
\ee
which, by the Stokes theorem, gives
\be
       X^{\mu \nu} = \frac{1}{2} \oint_{\Sigma_B} \dd s 
       \left ( X^\mu \frac{\p X^\nu}{\p s} - X^\nu \frac{\p X^\mu}{\p s}
       \right ) .
\lbl{5.1d}
\ee
Here $X^\mu (s)$ are embedding functions of the boundary loop $B$, $s$
being a parameter along the loop. Eq.\,(\ref{5.1d}) tells us that there is
a mapping
\be
        X^\mu (s)  \rightarrow X^{\mu \nu}
\lbl{5.1e}
\ee
from infinite dimensional objects $X^\mu (s)$, describing loops, into the finite
dimensional objects $X^{\mu \nu}$.

The above arrangement can describe two physically distinct situations:

\ \ (i) A loop $B$ can be a closed string. Then $X^{\mu \nu}$
are bivector coordinates associated with the closed string.

\ (ii) A surface $\Sigma$ can correspond to an open 2-brane whose boundary
is $B$. Then $X^{\mu \nu}$
are bivector coordinates associated with the open 2-brane.

Analogous setup holds for objects of arbitrary dimensions:
\be
X^{\mu_1 \mu_2 ... \mu_3} = \frac{1}{2} \int_{B_r} \dd \xi^{a_1 ...a_r}
\,  \p_{[a_1} X^{\mu_1} ... \p_{a_r]} X^{\mu_r} .
\lbl{5.1f}
\ee
    
Instead of the usual relativity, formulated
in spacetime in which the interval is
\be
{\rm{d}}s^2 \, = \,\,\eta _{\mu \nu \,} {\rm{d}}x^\mu  {\rm{d}}x^\nu  ,
\lbl{5.5}
\ee
one can consider the theory in which the interval is extended to
the space of $r$-volumes, called pandimensional continuum\,\ci{Pezzaglia} or
Clifford space\,\ci{CastroCspace,PavsicBook}:
\be
{\rm{d}}S^2 \, = \,G_{MN} \,{\rm{d}}x^M {\rm{d}}x^N .
\lbl{5.6}
\ee
Coordinates of Clifford space can be used to model extended
objects \ci{CastroCspace,PavsicArena,AuriliaQuenched,AuriliaFuzzy}.
They are a generalization of the concept of center of mass.
Instead of describing an extended object in ``full detail", we
can describe it in terms of the center of mass, area and
volume coordinates. In particular, the extended object can be a
fundamental string/brane.

{\it Dynamics.}  Taking also a time like parameter $\tau$, our object
can be described by 16 functions $X^M (\tau)$.
Let the action for an extended object described in terms of
the coordinates of Clifford space be
\be
I = \int \dd \tau \,(G _{MN} \dot X^M \dot X^N  )^{1/2} .
\lbl{5.7}
\ee
If $G_{MN}=\eta_{MN}$ is Minkowski metric, then the equations
of motion are
\be
\ddot X^M \, \equiv \,\,\frac{{\,{\rm{d}}^{\rm{2}} X^M }}{{{\rm{d}}\tau ^2 }}
\,\, = \,\,0 .
\lbl{5.8}
\ee
They hold for tensionless branes. For the branes with tension one has to replace
$\eta_{MN}$ with a generic metric $G_{MN}$ with non vanishing curvature.
Eq.\,(\ref{5.8}) then generalizes to the corresponding geodesic equation
\be
     \frac{1}{\sqrt{{\dot X}^2} }\, \frac{\dd}{\dd \tau}
     \left ( \frac{{\dot X}^M}{\sqrt{{\dot X}^2}}  \right )
     + \Gamma_{JK}^M \frac{{\dot X}^J {\dot X}^K}{{\dot X}^2} = 0 .
\lbl{5.9}
\ee     

As an example let us consider the Dirac membrane, described
by four embedding functions of three parameters
$\xi^a = (\tau, \vartheta, \varphi)$:
\be
X^\mu  (\xi ^a )\, = \,(X^0 ,\,\,r\,\sin \,\vartheta \,\cos \,\phi
 ,\,\,r\,\sin \,\vartheta 
\,\sin \,\phi ,\,\,r\,\cos \vartheta ) .
\lbl{5.9a}
\ee
The induced metric on the worldsheet swept by the membrane is
\be
  f_{ab} \, = \,\left( \begin{array}{l}
 \dot X^2 _0 \, - \,\dot r^2 \,\,\,\,\,\,\,0\,\,\,\,\,\,\,\,\,\,\,\,\,\,\,0 \\ 
 \,\,\,\,\,\,\,0\,\,\,\,\,\,\,\,\,\, - r^2 \,\,\,\,\,\,\,\,\,\,\,\,0 \\ 
 \,\,\,\,\,\,\,0\,\,\,\,\,\,\,\,\,\,\,\,0\,\,\, - r^2 \sin ^2 \,\vartheta \, \\ 
 \end{array} \right) .
\lbl{5.10}
\ee

The action is
\be
I = \int {{\rm{d}}\tau \,{\rm{d}}\vartheta \,{\rm{d}}\phi 
\,\sqrt {|f|} }  
= \int {{\rm{d}}\tau \,4\pi r^2 \sqrt {\dot X^2 _0  - \dot r^2 } } ,
\lbl{5.11}
\ee
where $\sqrt {|f|} \equiv  \sqrt {\,|\det f|}    = 
\sqrt {\dot X^2 _0  - \dot r^2 }  \, r^2 \sin \vartheta$.
Variation of the above action with respect to $r$ and $X^0 = X_0$ gives
the following equations of motion:
\bear
 &&\frac{{\rm{d}}}{{{\rm{d}}\tau }}\left( {\frac{{\dot r}}{{\sqrt {\dot X^2 _0 
 - \dot r^2 } }}} \right) + 
 \frac{{2\dot X^2 _0 }}{{r\sqrt {\dot X^2 _0  - \dot r^2 } }}\,\, 
= \,\,0 \lbl{5.12a} \\ 
 &&\frac{{\rm{d}}}{{{\rm{d}}\tau }}
 \left( {\frac{{r^2 \,\dot X_0 }}{{\sqrt {\dot X^2 _0  
- \dot r^2 } }}} \right)\,\, = \,\,0  . \lbl{5.12b} 
\ear 
If we now introduce the new variable according to eq.\,(\ref{5.1f})
\bear
 X^{123}  &=& \frac{1}{{3!}}\int {{\rm{d}}r\,{\rm{d}}\vartheta \,{\rm{d}}\phi } 
\,\,\partial _{[a} X^1 \partial _b X^2 \partial _{c]} X^3  = 
 \frac{{4\pi r^3 }}{3} \lbl{5.13a}\\  
  \dot X^{123}  &=& 4\pi r^2 \dot r \lbl{5.13b}\\  
\frac{{dX^{123} }}{{dS\,\,\,}} &=&
 \frac{{\dot X}^{123}}{4\pi r^2 \sqrt {\dot X^2 _0 - \dot r^2 } } 
 = \frac{{\dot r}}{{\sqrt {\dot X^2 _0  - \dot r^2 } }}  
\lbl{5.13}
\ear
where ${\rm{d}}S = 
\,\,{\rm{d}}\tau \,4\pi r^2 \sqrt {{\dot X}^2 _0  - {\dot r}^2}$,
then the equation of motion (\ref{5.12a}) becomes
\be
\frac{{{\rm{d}}^2 X^{123} }}{{{\rm{d}}S^2 \,\,\,\,}} + 
\frac{2}{{\,3X^{123} }}\left( {1 + \left( 
{\frac{{{\rm{d}}X^{123} }}{{{\rm{d}}S^{} 
\,\,\,\,}}} \right)^2 } \right) = 0 ,
\lbl{5.14}
\ee
whereas eq.\,(\ref{5.12b}), due to the reparametrization invariance
of our action (\ref{5.11}), is redundant.

The  equation of motion (\ref{5.14}) can be considered as the geodesic
equation (\ref{5.9}) derived from the $C$-space action (\ref{5.7})
for the case of a subspace described by two coordinates
$X^M = (X^0, X^{123})$ with the metric
\be
G_{MN}  = \left( \begin{array}{l}
 C\tilde X^{4/3} \,\,\,\,0 \\ 
 0\,\,\,\,\,\,\,\,\,\,\, - 1 \\ 
 \end{array} \right) ,
\lbl{5.15}
\ee
where $C$ is a constant, and where we have denoted ${\tilde X} \equiv X^{123}$.
Namely, if we insert the particular metric (\ref{5.15}) into the
equation of geodesic (\ref{5.9}), then we obtain eq.\,(\ref{5.14}).

We can show that the above $C$-space description is equivalent
to the Dirac membrane by directly comparing the actions.
In the 2-dimensional subspace with coordinates $X^M = (X^0,X^{123})$,
$X^{123} \equiv {\tl X}$, and the metric (\ref{5.15}) we have the
following line element
\be
 {\rm{d}}S^2  = G_{00} ({\rm{d}}X^0 )^2  +
  G_{\tilde X\tilde X} {\rm{d}}\tilde X^2  
   = C\tilde X^{4/3} ({\rm{d}}X^0 )^2 \, - {\rm{ d}}\tilde X^2 
\lbl{5.16}
\ee
Using
\bear
 &&\tilde X = \frac{{4\pi r^3 }}{3}\,\,,\,\,\,\,\,\,\,\,\,\,\,{\rm{d}}\tilde X\, 
 = 4\pi r^2 {\rm{d}}r \lbl{5.17a}\\ 
&& \tilde X^{4/3}  = \left( {\frac{{4\pi }}{3}} \right)^{4/3} r^4  \lbl{5.17b}\\ 
 && C\left( {\frac{{4\pi }}{3}} \right)^{4/3} = (4\pi )^2 
\lbl{5.17}
\ear
we have
\be
{\rm{d}}S^2  = (4\pi r^2 )^2 \left( ({{\rm{d}} X^0 )^2  - {\rm{d}}r^2 } \right) .
\lbl{5.18}
\ee
Inserting the latter line element into the action
\be
I[X^M ] = \int {{\rm{d}}S}  = \int_{}^{} 
{d\tau \,(G_{MN} \dot X^M \dot X^N } )^{1/2} \, ,
\lbl{5.19}
\ee
we obtain
\be
I = \int {{\rm{d}}\tau } \,(4\pi r^2 )^2 \sqrt {(\dot X^0 )^2  - \dot r^2 } .
\lbl{5.20}
\ee
which is the action for the Dirac membrane.

The above example explicitly shows why description of branes with
non vanishing tension requires non trivial metric of the brane
configuration space $C$. The $C$-space metric (\ref{5.15}) that gives
the usual membrane action (\ref{5.20})  is just one particular case. Other
more general $C$-space metrics are possible in this theory.
Such higher dimensional configuration space, associated with branes, enables
unification of fundamental interactions \` a la
Kaluza-Klein\,\ci{PavsicKaluza}.
For alternative, although related approaches see \ci{CliffKaluzaAlternative}.

\section{Conclusion}

We have considered a theory in which spacetime is replaced by a larger space,
namely the configuration space ${\cal C}$,  associated with a system
under consideration. In this picture a 4-dimensional spacetime is just
a subspace of ${\cal C}$, associated with the degrees of freedom of a single point
particle. For strings/branes the configuration space
is infinite dimensional, but it can be reduced to a corresponding
finite dimensional space, the so called Clifford space $C$.
Since configuration space has extra dimensions,
its metric provides a description of additional interactions, beside the
4-dimensional gravity, just as in Kaluza-Klein theories.
In this theory there is no need for extra dimensions of {\it spacetime}, because
the latter space is a subspace of the {\it configuration space} ${\cal C}$,
and all dimensions of ${\cal C}$ are physical.
Therefore, there is no need for a compactification of the extra dimensions of 
${\cal C}$. The additional interactions, besides the 4D gravity, can occur in macro
physics and in micro physics.
If we consider macroscopic systems, e.g., at the levels from galaxies
to the Universe, then we have a modification of gravity.
If we consider elementary particles, then we obtain
a description of gauge interactions. For larger  values of
$n$ (number of particles), one would expect
gauge groups of larger and larger rank, but only in general. In particular,
the configuration space metric can be of such block diagonal form
(one block for each particle)  that the gauge group of the interactions between
fundamental particles remain the same. How this works in detail we
postpone to future investigations.

The notion of configuration space is ubiquitous in standard physical thought,
and there are the subtle differences between how it and the usual space are
are usually treated. But according to the view described in this paper, there
are no such subtle differences: Basically there is the space of all possible
matter configurations. If one assumes that positions of all particles
are fixed and only the position of one particle is
variable, then one has the space of all possible positions of the single
particle. This is just the 4-dimensional spacetime. But the latter space
is not the whole story, since it is only  a subspace of a more general
configuration space. What the standard physical thought has missed so
far is to fully employ this more general space in the formulations of
physical theories.

Very relevant for understanding the role of configuration space
and related concepts, that go far beyond the standard physical thought,
are works by Barbour\,\ci{Barbour} and Anderson\,\ci{AndersonConfig} .
In those works, matter configurations have been described in the intrinsic
terms, without recourse to an embedding space or spacetime, therefore usage
of coordinates has been avoided. Also in our approach every possible matter
configuration is represented as a point in a configuration space. But,
following general relativity, we adopt the view that to points of ${\cal C}$
we can assign arbitrary coordinates. And, like ``house numbers'', the
set of coordinates assigned to a point in ${\cal C}$ can be changed into
another set of coordinates. So our approach retains the crucial
feature of general relativity, such as diffeomorphism invariance and
background independence.

\end{document}